%% file: main.tex
\documentclass{article}

\usepackage[dblblindworkshop, final]{neurips_2026_vericode}
\usepackage[utf8]{inputenc}
\usepackage[T1]{fontenc}
\usepackage{hyperref}
\usepackage{url}
\usepackage{booktabs}
\usepackage{amsfonts}
\usepackage{amsmath}
\usepackage{microtype}
\usepackage{xcolor}
\usepackage{array}
\usepackage{tabularx}
\usepackage{graphicx}
\usepackage{listings}

\lstdefinestyle{appendixblock}{
  basicstyle=\fontencoding{T1}\ttfamily\footnotesize,
  breaklines=true,
  breakatwhitespace=false,
  columns=fullflexible,
  keepspaces=true,
  showstringspaces=false,
  frame=single,
  framerule=0.3pt,
  rulecolor=\color{black!35},
  backgroundcolor=\color{black!2},
  xleftmargin=0.4em,
  xrightmargin=0.4em,
  aboveskip=0.6em,
  belowskip=0.6em,
  escapeinside={(*@}{@*)}
}

\title{Eliciting Specifications: When Does Interactive Clarification Help?}
\workshoptitle{AI for Verifiable Coding}
\author{%
  Mowafak Allaham\thanks{Corresponding author: \texttt{mo@emergence.ai}} \quad
  Aditya Vempaty \quad Ashish Jagmohan \quad Prasenjit Dey \quad Ravi Kokku \\
  Emergence AI
}

\begin{document}
\maketitle

\begin{abstract}
LLMs can generate Lean specifications from task descriptions, but those descriptions are often underspecified or ambiguous and may omit verifier-relevant edge cases, leading to imperfect formal specifications. We study whether an elicitation agent (EA) can improve verifier-checked Lean specifications by seeking clarification from a reference-grounded simulated developer before formalization. The agent decomposes the task, identifies unresolved decisions, asks bounded clarification questions, and uses the answers to write a refined natural-language specification that an LLM formalizer then translates into Lean. On 189 Verina Lean 4 specification-generation tasks \citep{ye2026verina}, we compare single-shot Lean specification generation with elicitation under question budgets of $k=3$ and $k=10$, allowing at most three or ten EA questions per task, across same-model and cross-model configurations. The same-model Claude Sonnet 4.6 pipeline has negative strict pass@1 point estimates at both budgets, showing that elicitation is not a uniform accuracy booster. In contrast, the complete Sonnet 4.6-to-GPT-5.6 and Sonnet 4.6-to-Gemini 3.8 Flash pipeline configurations have positive strict point estimates. The largest configuration-level strict result reaches 65.6\%, a +6.3-point change (+12 tasks), in the Sonnet 4.6-to-GPT-5.6 pipeline. Six of eight conditions have positive compile-conditional point estimates, with a maximum cross-model difference of +4.9 percentage points among compiling outputs. Treatment point estimates also differ across the basic and advanced benchmark splits; the largest advanced-task change occurs in the Sonnet 4.6-to-GPT-5.6 configuration at $k=10$, from 42.0\% to 50.6\%. Failure analysis identifies postconditions as the component with the largest degradations. Together, these findings suggest that future autoformalization systems should treat elicited natural-language specifications as intermediate artifacts and optimize jointly for semantic correctness among compiling outputs and reliable translation of clarified information into Lean.
\end{abstract}

\input{sections/01_introduction_rqs}
\input{sections/06_related_work}
\input{sections/02_methods}

\input{sections/04_main_results}

\input{sections/05_mechanism_failure_analysis}
\input{sections/07_discussion}
\input{sections/08_limitations}
\input{sections/09_conclusion}

\bibliographystyle{plainnat}
\bibliography{references}

\clearpage
\appendix
\input{sections/10_appendix}

\end{document}

%% file: sections/01_introduction_rqs.tex
\section{Introduction}

\label{sec:intro}

Formal verification can make generated software accountable to an explicit specification, but that guarantee is only as meaningful as the specification being checked \citep{ye2026verina}. Task descriptions may leave boundary cases, representation choices, invariants, and intended failure behavior unstated or ambiguous \citep{vijayvargiya2025ambigswe,berry2003ambiguity}. Autoformalization in Lean must therefore solve two linked problems: resolve relevant decisions left open by the description, and express the resulting behavior as preconditions and postconditions that compile and satisfy semantic checks \citep{wu2022autoformalization,ye2026verina}.

The standard single-shot SpecGen setting combines these problems into one step \citep{ye2026verina}. An LLM receives the original task description and function signature, then directly writes a Lean specification. This setup is reproducible, but it forces the model to silently resolve ambiguities before committing them to a formal statement. If a prompt does not specify how ties should be handled, which values must be ruled out, or what property an output must witness, the model must guess. Such decisions can instead be clarified before they become part of a formal contract \citep{goguen1993techniques,vijayvargiya2025ambigswe}.

Dialogue-based elicitation provides a natural way to expose these decisions. We use \emph{interactive clarification} to denote this question-and-answer process. In our setting, an elicitation agent decomposes the task, identifies unresolved choices, conducts a bounded clarification dialogue, and writes a refined natural-language specification before Lean formalization \citep{wang2026turningintent,chen2026specfirst}. We refer to the downstream LLM that converts either the original description or the refined natural-language specification into Lean as the LLM formalizer (Section~\ref{sec:ea}). This design separates the production of a clarified natural-language handoff from the final Lean transcription step: the elicitation agent may surface useful distinctions before formalization, but end-to-end success still depends on whether the LLM formalizer preserves those distinctions in compilable and semantically correct Lean.

This paper evaluates dialogue-based specification elicitation on the full Verina benchmark of 189 Lean 4 specification-generation tasks, split into 108 basic and 81 advanced problems \citep{ye2026verina}. Each task is evaluated under a single-shot baseline and under elicitation with question budgets of $k=3$ and $k=10$, permitting at most three or ten EA questions per task. The experiments include a same-model setting and cross-model settings. In each cross-model treatment run, Sonnet 4.6 powers the elicitation agent and simulated developer, while GPT-5.6, Gemini 3.8 Flash, or Claude Sonnet 5 serves as the LLM formalizer. The dialogue and refined handoff are sampled independently in each configuration; the cross-model comparisons therefore characterize complete pipeline configurations rather than a formalizer-only intervention and do not rank formalizers on a shared elicited specification. We report strict pass@1 as the primary end-to-end metric and compile-conditional pass@1 as a diagnostic for correctness among generated specifications that compile (Section~\ref{sec:methods}).

Here, dialogue-based elicitation denotes bounded interaction with a reference-grounded simulated developer. The experiment measures whether an EA can obtain and transmit useful benchmark-grounded information through that controlled interaction; it does not evaluate elicitation from human users.

The study addresses two research questions:

\textbf{RQ1: Does dialogue-based elicitation improve Lean specification generation relative to single-shot generation from the original task descriptions?} We answer this question using strict pass@1 and compile-conditional pass@1 across the full Verina task set and two clarification budgets (Table~\ref{tab:main-results}).

\textbf{RQ2: When do elicited natural-language specifications translate into end-to-end autoformalization gains?} We answer this question by comparing task difficulty, question budget, and complete same-model and cross-model configurations, including configurations with different receiving LLM formalizers (Table~\ref{tab:main-results} and Appendix~\ref{app:secondary-results}).

The results show that elicitation is conditional rather than uniformly beneficial. In the same-model deployment, Claude Sonnet 4.6 loses solved tasks when it both elicits and formalizes, with losses concentrated in postconditions. The complete Sonnet 4.6-to-GPT-5.6 and Sonnet 4.6-to-Gemini 3.8 Flash pipelines have positive strict point estimates. Claude Sonnet 5 shows that stronger baseline performance alone does not guarantee gains from elicitation: despite the strongest single-shot baseline, the Sonnet 4.6-to-Claude Sonnet 5 configuration regresses at $k=10$, again through postcondition degradation. Compile-conditional pass@1 has positive point estimates in six of eight primary treatment conditions, a pattern consistent with the refined natural-language handoff being more useful among outputs that the LLM formalizer can express as valid Lean. These findings characterize the observed value of reference-grounded clarification across complete model configurations; because the realized dialogues differ across configurations, they do not isolate the formalizer as the sole cause of cross-model differences or rank formalizers on a common handoff.

The paper makes three contributions. First, it evaluates EA-mediated, reference-grounded clarification on the full Verina benchmark, using two question budgets and multiple LLM formalizers. Second, it separates strict pass@1 from compile-conditional pass@1, distinguishing end-to-end outcomes from correctness rates among compiling specifications. Third, it shows that observed treatment differences vary across model configurations: Sonnet 4.6 to GPT-5.6 and Sonnet 4.6 to Gemini 3.8 Flash produce positive point estimates, while the Claude Sonnet 5 result shows that stronger baseline performance alone does not guarantee an elicitation gain and identifies postcondition degradation as the largest negative component.

%% file: sections/06_related_work.tex
\section{Related Work}
\label{sec:related}

\textbf{Autoformalization and verifiable code generation.} Autoformalization studies how models translate informal mathematical or programming intent into formal statements, while verifiable code generation treats code, specifications, and proofs as linked artifacts whose correctness can be evaluated separately \citep{wu2022autoformalization,ye2026verina}. Verina provides the benchmark setting for this study because it exposes the specific failure mode we want to measure: an informal task description must become a Lean pre/postcondition artifact that compiles and satisfies separate soundness and completeness checks \citep{ye2026verina}. This differs from benchmarks that evaluate generated code, free-form answers, or interactive task completion without a proof-assistant checker in the loop. Verina is therefore not only a source of tasks, but the evaluation interface that lets us distinguish compilation failures from semantic failures after elicitation.

\textbf{Pre- and post-formalization refinement.} VeriSpecGen and our method intervene at different stages of specification synthesis \citep{ye2026verispecgen}. VeriSpecGen decomposes requirements, generates an initial Lean specification, and uses requirement-targeted tests and verifier feedback to localize and repair incorrect clauses. Our study intervenes before formalization: an elicitation agent asks a reference-grounded simulated developer questions, then passes the resulting natural-language description to a one-shot LLM formalizer. The approaches are therefore complementary. Raw performance is not directly comparable because VeriSpecGen uses iterative test- and verifier-guided repair, whereas our experiment measures the effect of information acquired through pre-formalization clarification.

\textbf{Ambiguity and intent elicitation.} Research on elicitation and specification has long treated ambiguity as a process problem. Stakeholders often leave assumptions tacit, so analysts must elicit goals, constraints, and contextual commitments before writing a specification \citep{goguen1993techniques}. Ambiguity in software specifications can arise from underspecified references, scope, quantification, and ordinary-language terms that admit multiple formal interpretations \citep{berry2003ambiguity}. Recent software-agent benchmarks bring this concern into LLM coding workflows by testing whether agents detect missing information, ask useful clarification questions, and use the answers productively \citep{vijayvargiya2025ambigswe,pu2026lhaw,trinh2026hilbench,raghavendra2026sweinteract}. These benchmarks are important for studying interaction, but they generally do not ask whether clarified task information survives translation into verifier-checked Lean preconditions and postconditions. That gap motivates using Verina rather than a general software-interaction benchmark.

\textbf{Human-centered verification.} Human-centered verification emphasizes mixed-initiative support, explanations, trust calibration, and accessibility for developers without formal-methods expertise \citep{trinh2026hilbench,raghavendra2026sweinteract,chang2026criticality}. That literature motivates clarification before formalization, but the present study does not evaluate human intent elicitation, an interface, or a human-subject workflow. It instead evaluates a controlled EA interaction with a reference-grounded simulated developer.

\textbf{Interactive specification workflows.} Specification-centered agent work argues that intent discovery should occur before implementation or synthesis \citep{wang2026turningintent,chen2026specfirst,hamblin2026specbench}. These systems motivate the idea that a model should clarify a task before producing a downstream artifact, but their evaluation targets differ from ours. Some emphasize structured requirements, executable specifications, critique of incomplete specifications, or downstream program behavior. Our study focuses on a narrower formal interface: the elicitation agent writes a refined natural-language handoff, and a separate LLM formalizer must translate that handoff into Lean predicates that are checked for compilation, soundness, and completeness. The novelty is therefore not the existence of clarification dialogue by itself, but the measurement of when a clarified handoff helps or hurts in a verifier-scored autoformalization pipeline.

\textbf{Model guidance and transfer.} Advisor Models show that instance-specific natural-language guidance can transfer across receiving models \citep{asawa2025advisor}. Our cross-model design is analogous at the configuration level: the elicitation agent produces the clarified description, and the LLM formalizer consumes it. The proof-assistant setting adds a stricter bottleneck. Guidance that is useful in natural language must still be translated into valid Lean predicates, and failures in that translation receive no strict credit. This is the central gap addressed by our study: we evaluate not only whether elicitation changes the information available to the model, but when that information survives formalization into verifiable Lean specifications.

%% file: sections/02_methods.tex
\section{Methods}
\label{sec:methods}

\input{sections/02_elicitation_agent}
\input{sections/03_experimental_design_metrics}

%% file: sections/02_elicitation_agent.tex
\subsection{Elicitation}

\label{sec:ea}

The elicitation agent (EA) is a structured preprocessing stage for specification generation. Given the original Verina task description and Lean function signature, the EA identifies underspecified choices, asks a bounded number of clarification questions, and writes a refined natural-language specification before any Lean code is generated. This design follows work on elicitation and ambiguity that treats ambiguity as something to expose before committing to a specification, and human-AI work showing that structured intermediate reasoning can help users inspect and repair assumptions \citep{goguen1993techniques,berry2003ambiguity,chang2026criticality}. The downstream model that converts either the original description or the refined description into Lean is the LLM formalizer.

Appendix Figure~\ref{fig:ea-architecture} summarizes the treatment pipeline, and Appendix~\ref{app:figure1-example} provides the corresponding raw trace, including the EA's decomposition, clarification dialogue, refined handoff, and generated Lean specification. The EA first decomposes the task into a structured frame covering the task purpose, assumptions, relevant concepts, evidence from the prompt, and unresolved decisions. It also builds a small decision space of possible interpretations, such as input restrictions, edge cases, ordering constraints, or output relationships. The agent then selects unresolved decisions to ask about within the experiment's question budget. In this paper, the budgets are $k=3$ and $k=10$, where $k$ is the maximum number of clarification questions.

The dialogue partner is a reference-grounded simulated developer with access to the reference Lean bundle and tests. It is intentionally permitted to provide requested reference-grounded semantic information. Neither the EA nor the LLM formalizer accesses the reference directly: such information can reach the formalizer only when the EA asks a relevant question, the simulated developer answers it, and the EA carries that answer into the refined handoff. Because the target artifact is a Lean specification, the EA may ask formalization-aware questions about quantifier structure, boundary cases, representation choices, and postcondition form. The simulated developer is constrained to answer only the question asked and not to provide Lean code, verbatim reference preconditions or postconditions, or symbolic formulas. Natural-language answers may nevertheless convey reference-specific contract structure; this is the intended information channel rather than direct artifact access. The experiment therefore measures whether a downstream LLM formalizer can benefit from bounded, reference-grounded clarification before Lean generation; it does not establish that the same gains would hold with human developers or non-reference-grounded answers. This distinction mirrors recent interactive-agent benchmarks that separate asking for missing information from directly observing the desired artifact \citep{vijayvargiya2025ambigswe,trinh2026hilbench,raghavendra2026sweinteract}.

After the bounded dialogue, the EA writes a final natural-language specification. The refinement prompt asks for the intended precondition, intended postcondition, edge-case clarifications, and any corrections to earlier assumptions. The handoff is not target Lean code, but it can include formalization-relevant distinctions that the dialogue has surfaced. The LLM formalizer then receives this refined description through the same Verina SpecGen interface used by the baseline. This separation is central to the study: the EA produces a clarified natural-language handoff, while the LLM formalizer determines whether that handoff survives as a compiling and correct Lean artifact.

%% file: sections/03_experimental_design_metrics.tex
\subsection{Experimental Design}

The experiment evaluates whether bounded clarification dialogue improves Lean 4 specification generation on the full Verina SpecGen dataset \citep{ye2026verina}. The benchmark contains 189 tasks, split into 108 basic and 81 advanced tasks. Each task provides an informal description, a Lean function signature, reference Lean code, tests, and a difficulty label. The generation target is fixed across all conditions: for each task, the system produces one candidate Verina specification artifact containing imports, precondition definitions, precondition, postcondition definitions, and postcondition. We score generated artifacts with Verina's benchmark evaluation scripts and will release the prompts used for the baseline LLM formalizer, elicitation agent, simulated developer, and refinement step.

The simulated developer is implemented as a separate LLM call, powered by Claude Sonnet 4.6 in all four primary configurations. For each answer, its system prompt contains the complete task-specific Verina reference Lean file---including the reference implementation and intended pre- and postconditions---together with the benchmark's positive and rejection test inputs. These fields are loaded directly from the task record; the simulated developer performs no external retrieval or search. The current EA question is supplied as the sole user message. The prompt requires a direct natural-language reply of at most three sentences, limits the answer to the question asked, and prohibits Lean syntax, tactics, code blocks, and verbatim reference predicates. The EA and LLM formalizer never receive the reference bundle directly; they can receive its semantic content only through answers elicited by the EA.

Each task is evaluated under baseline and elicitation conditions, as shown in Figure~\ref{fig:experimental-design}. In the baseline condition, the LLM formalizer receives the original Verina task description and function template in a single shot and produces one Lean specification candidate. In the elicitation condition, the EA receives the same initial task, conducts a bounded clarification dialogue with the simulated developer, writes a refined natural-language specification, and passes that specification to the LLM formalizer prompt used by the baseline. The LLM formalizer again produces one candidate, which is scored by the same Verina evaluator. The LLM formalizer and the evaluator are therefore held fixed within each baseline--treatment comparison; the treatment changes the natural-language handoff supplied before formalization.

\begin{center}
\refstepcounter{figure}\label{fig:experimental-design}
\includegraphics[width=0.98\linewidth]{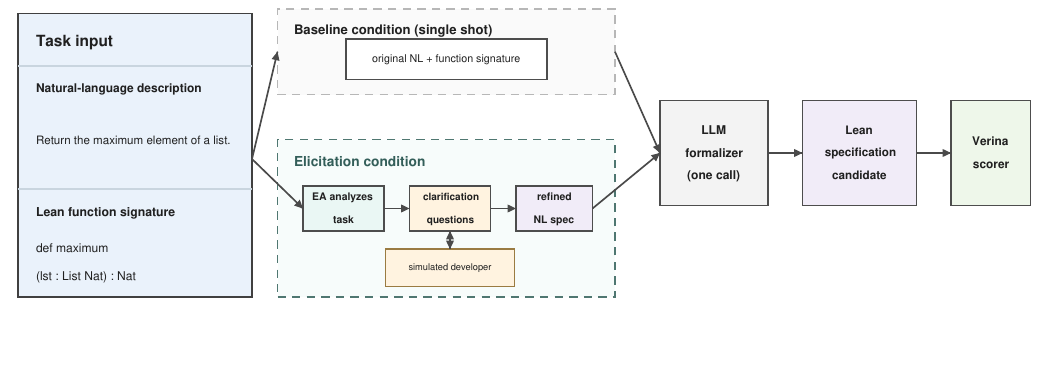}
\vspace{0.2em}
\begin{minipage}{0.94\linewidth}
\textbf{Figure~\thefigure:} Matched baseline and elicitation conditions. Both arms begin with the same natural-language task description and Lean function signature. The baseline sends this input to the LLM formalizer in a single shot; the elicitation arm first produces a refined natural-language specification through EA analysis and bounded clarification. Each arm generates one Lean specification candidate, scored by the same Verina evaluator.
\end{minipage}
\end{center}

The design follows the paper's two research questions. To test whether elicitation helps, we compare single-shot generation with question budgets $k=3$ and $k=10$, where $k$ is the maximum number of EA questions per task. The EA can stop when it selects no further question. Across the four primary configurations, the mean realized count ranges from 2.58 to 2.67 questions at $k=3$ and from 6.93 to 7.17 questions at $k=10$. The corresponding budget-exhaustion rates are 76.2--80.4\% and 20.6--25.4\%, respectively (Appendix Table~\ref{tab:realized-questions}). Thus, the comparison estimates the effect of offering a larger question allowance under the EA's stopping policy; it does not isolate an effect of realized dialogue length. To test when elicitation helps, we compare configurations with different receiving LLM formalizers while keeping the task set, generation target, scorer, and EA and simulated-developer model identities fixed. In the same-model arm, Claude Sonnet 4.6 fills all three model roles. In the cross-model arms, Claude Sonnet 4.6 powers the EA and simulated developer, while the LLM formalizer is GPT-5.6, Gemini 3.8 Flash, or Claude Sonnet 5. The EA--simulated-developer interaction is sampled separately in each configuration, so the realized dialogues and refined descriptions are not held fixed across receiving formalizers. These comparisons evaluate complete model configurations, are not formalizer-only ablations, and cannot rank formalizers on a shared elicited specification.

\subsection{Metrics}

A generated specification must pass two gates before it receives credit. The first gate is syntactic: both the precondition and postcondition must compile as valid Lean. The second gate is semantic: the compiled specification must pass Verina's soundness and completeness checks. The two reported metrics use these gates differently.

\textbf{Strict pass@1.} Strict pass@1 is the primary end-to-end metric. It counts specifications that pass both gates and divides by all 189 tasks. A noncompiling specification is therefore a failure in the same way that a compiling but logically incorrect specification is a failure. The metric measures whether the available natural-language information, either original or elicited, has been successfully translated into a valid and correct Lean specification, but it does not by itself distinguish compilation failures from semantic failures.

\textbf{Compile-conditional pass@1.} Compile-conditional pass@1 uses the same sound-and-complete success criterion, but divides only by generated specifications that pass the compilation gate. It discards noncompiling outputs and asks: among specifications the LLM formalizer managed to express as valid Lean, how many were also semantically correct? It is a diagnostic metric, not a replacement for strict pass@1. When the observed compile-conditional pass proportion is higher while strict pass@1 is lower, the compiling treatment outputs have a higher semantic pass proportion, but baseline and treatment condition on different subsets. The comparison therefore does not isolate a causal improvement in correctness.

We also report three secondary breakdowns. First, we split strict pass@1 by Verina difficulty label to compare basic and advanced tasks. Second, we count task-level gains and regressions, where a gain is a task that fails in the baseline and passes after elicitation, and a regression is the reverse transition. Third, we decompose strict pass@1 into precondition soundness, precondition completeness, postcondition soundness, and postcondition completeness. These breakdowns identify the task strata and specification components associated with changes in the reported metrics.

Because baseline and elicitation outcomes are observed on the same Verina tasks, we estimate uncertainty over paired task-level changes. For strict pass@1, we report exact sign-test p-values, equivalent to McNemar tests over discordant task pairs, and paired bootstrap 95\% confidence intervals for the treatment-minus-baseline delta. For compile-conditional pass@1, the compiling set can change between baseline and treatment, so McNemar testing is not appropriate; we instead report task-level bootstrap confidence intervals as diagnostic uncertainty estimates. These intervals quantify task-level uncertainty, not within-model sampling variance from repeated generations.

%% file: sections/04_main_results.tex
\newcommand{\tri}{\raisebox{-0.05em}{\includegraphics[height=0.72em]{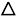}}}

\section{Results}

\label{sec:results}

The results compare single-shot Lean specification generation with refined natural-language specifications produced by the elicitation agent on the full Verina benchmark, using 189 tasks: 108 basic and 81 advanced. Treatment permits at most three or ten EA questions per task ($k=3$ and $k=10$) before the LLM formalizer produces the final Lean 4 specification. Following Section~\ref{sec:methods}, strict pass@1 is the primary end-to-end metric. A generated specification receives strict credit only when both its precondition and postcondition compile and pass Verina's soundness and completeness checks.

\begin{center}
\refstepcounter{table}\label{tab:main-results}
\begin{minipage}{\textwidth}
\textbf{Table~\thetable:} Overall strict pass@1 on the full Verina SpecGen set, n=189. Baseline is single-shot Lean specification generation without the elicitation agent. Elicitation columns report maximum EA-question budgets of $k=3$ and $k=10$ per task. Percentages are passing tasks out of 189; \tri{} is shown in percentage points and raw task count. Compile-conditional \tri{} reports the corresponding percentage-point change among generated specifications that compile.
\begin{center}
\resizebox{\textwidth}{!}{%
\begin{tabular}{lcccccc}
\toprule
\multicolumn{1}{c}{Configuration} & \begin{tabular}{@{}c@{}}Baseline\\(single-shot)\end{tabular} & \begin{tabular}{@{}c@{}}Maximum budget\\($k=3$)\end{tabular} & \begin{tabular}{@{}c@{}}\tri{}\\pp (tasks)\end{tabular} & \begin{tabular}{@{}c@{}}Maximum budget\\($k=10$)\end{tabular} & \begin{tabular}{@{}c@{}}\tri{}\\pp (tasks)\end{tabular} & \begin{tabular}{@{}c@{}}Compile-cond. \tri{} (pp)\\$k=3$ / $k=10$\end{tabular} \\
\midrule
Sonnet 4.6 same-model & 60.3\% & 57.7\% & -2.6 pp (-5) & 56.1\% & -4.2 pp (-8) & +3.8 / +2.7 \\
Sonnet 4.6 to GPT-5.6 cross-model & 59.3\% & \textbf{65.6\%} & \textbf{+6.3 pp (+12)} & \textbf{65.6\%} & \textbf{+6.3 pp (+12)} & +4.6 / +3.8 \\
Sonnet 4.6 to Gemini 3.8 Flash cross-model & 59.8\% & 64.6\% & +4.8 pp (+9) & 64.6\% & +4.8 pp (+9) & \textbf{+4.9} / \textbf{+4.9} \\
Sonnet 4.6 to Claude Sonnet 5 cross-model & \textbf{63.5\%} & 64.0\% & +0.5 pp (+1) & 60.3\% & -3.2 pp (-6) & -1.9 / -2.3 \\
\bottomrule
\end{tabular}%
}
\end{center}
\end{minipage}
\end{center}
\textbf{Strict pass@1.} The single-shot baselines in Table~\ref{tab:main-results} establish the single-shot performance of the evaluated LLM formalizers in our experimental setup. Baseline strict pass@1 ranges from 59.3\% to 63.5\%, so the evaluated LLM formalizers solve a majority of Verina specification-generation tasks from the original task description alone, but still fail on roughly one-third or more of the benchmark under strict sound-and-complete scoring.

Elicitation changes this baseline performance in a configuration-dependent way, as summarized in Table~\ref{tab:main-results}. With Sonnet 4.6 in all roles, elicitation reduces strict pass@1 by 5 tasks at $k=3$ and 8 tasks at $k=10$. Two independently sampled cross-model configurations have positive strict point estimates: Sonnet 4.6 to GPT-5.6 gains 12 tasks at both budgets, while Sonnet 4.6 to Gemini 3.8 Flash gains 9 tasks at both budgets. In the Sonnet 4.6-to-GPT-5.6 configuration, the identical aggregate strict pass@1 at $k=3$ and $k=10$ masks a difficulty-level shift: the larger budget loses four basic-task passes but gains four advanced-task passes, leaving the total fixed at 124/189. Claude Sonnet 5 is the counterexample: its complete configuration gains 1 task at $k=3$ but loses 6 tasks at $k=10$. Because the EA--simulated-developer interaction was rerun in each configuration, these differences characterize complete pipelines rather than the isolated effect of changing the LLM formalizer; they do not rank formalizers on a shared elicited specification.

Paired uncertainty estimates support a conservative interpretation (Appendix Table~\ref{tab:strict-uncertainty}). The Sonnet 4.6-to-GPT-5.6 configurations provide the clearest strict-pass signal, with +6.3 percentage-point changes and bootstrap intervals whose lower bounds are approximately zero. The Sonnet 4.6-to-Gemini 3.8 Flash configurations have positive point estimates at both budgets, but their intervals cross zero. The Sonnet 4.6 same-model and Sonnet 4.6-to-Claude Sonnet 5 results should therefore be read as possible configuration-specific failure patterns rather than statistically conclusive aggregate regressions.

\textbf{Compile-conditional pass@1.} Compile-conditional pass@1 separates the two gates that strict pass@1 collapses. Strict pass@1 asks whether a task passes both compilation and semantic checking out of all 189 tasks. Compile-conditional pass@1 first conditions on the compilation gate and then asks whether the valid Lean specification is sound and complete. This distinction is necessary because a generated specification can compile while still expressing the wrong behavior.

The compile-conditional deltas in Table~\ref{tab:main-results} describe semantic success among outputs that compile. Six of eight treatment conditions have positive point estimates on this metric. Sonnet 4.6 same-model is the clearest split between the two metrics: strict pass@1 decreases, while the observed compile-conditional proportion increases by 3.8 percentage points at $k=3$ and 2.7 points at $k=10$. Because conditioning selects different compiling subsets in baseline and treatment, these differences are diagnostic rather than causal estimates of improved correctness.

The Sonnet 4.6-to-GPT-5.6 and Sonnet 4.6-to-Gemini 3.8 Flash configurations show positive compile-conditional point estimates: +4.6 and +3.8 points for the former, and +4.9 points at both budgets for the latter. The Sonnet 4.6-to-Claude Sonnet 5 configuration has negative point estimates of -1.9 points at $k=3$ and -2.3 points at $k=10$; Section~\ref{sec:mechanism} associates this pattern with postcondition failures. The bootstrap intervals for compile-conditional deltas are wide and cross zero in all conditions (Appendix Table~\ref{tab:compile-uncertainty}), so this metric is best interpreted as a diagnostic description rather than evidence that the intermediate natural-language specification improved as a standalone artifact.

\begin{center}
\refstepcounter{table}\label{tab:difficulty-baselines}
\begin{minipage}{\textwidth}
\textbf{Table~\thetable:} Strict pass@1 by Verina difficulty label. Baseline is single-shot Lean specification generation without the elicitation agent. Elicitation columns report maximum EA-question budgets of $k=3$ and $k=10$ per task. Percentages use the difficulty denominator, n=108 for basic and n=81 for advanced. \tri{} is shown in percentage points and raw task count.
\begin{center}
\resizebox{\textwidth}{!}{%
\begin{tabular}{llccccc}
\toprule
\multicolumn{1}{c}{Configuration} & \multicolumn{1}{c}{Difficulty} & \begin{tabular}{@{}c@{}}Baseline\\(single-shot)\end{tabular} & \begin{tabular}{@{}c@{}}Maximum budget\\($k=3$)\end{tabular} & \begin{tabular}{@{}c@{}}\tri{}\\pp (tasks)\end{tabular} & \begin{tabular}{@{}c@{}}Maximum budget\\($k=10$)\end{tabular} & \begin{tabular}{@{}c@{}}\tri{}\\pp (tasks)\end{tabular} \\
\midrule
Sonnet 4.6 same-model & Basic & 73.1\% & 65.7\% & -7.4 pp (-8) & 64.8\% & -8.3 pp (-9) \\
Sonnet 4.6 same-model & Advanced & 43.2\% & 46.9\% & +3.7 pp (+3) & 44.4\% & +1.2 pp (+1) \\
Sonnet 4.6 to GPT-5.6 cross-model & Basic & 72.2\% & \textbf{80.6\%} & \textbf{+8.3 pp (+9)} & 76.9\% & +4.6 pp (+5) \\
Sonnet 4.6 to GPT-5.6 cross-model & Advanced & 42.0\% & 45.7\% & +3.7 pp (+3) & \textbf{50.6\%} & \textbf{+8.6 pp (+7)} \\
Sonnet 4.6 to Gemini 3.8 Flash cross-model & Basic & \textbf{74.1\%} & 77.8\% & +3.7 pp (+4) & 79.6\% & +5.6 pp (+6) \\
Sonnet 4.6 to Gemini 3.8 Flash cross-model & Advanced & 40.7\% & 46.9\% & +6.2 pp (+5) & 44.4\% & +3.7 pp (+3) \\
Sonnet 4.6 to Claude Sonnet 5 cross-model & Basic & 72.2\% & 72.2\% & +0.0 pp (+0) & 75.9\% & +3.7 pp (+4) \\
Sonnet 4.6 to Claude Sonnet 5 cross-model & Advanced & \textbf{51.9\%} & \textbf{53.1\%} & +1.2 pp (+1) & 39.5\% & -12.3 pp (-10) \\
\bottomrule
\end{tabular}%
}
\end{center}
\end{minipage}
\end{center}

\textbf{Difficulty stratification.} Single-shot performance is higher on basic tasks (72.2--74.1\%) than on advanced tasks (40.7--51.9\%). The largest basic-task estimate is the Sonnet 4.6-to-GPT-5.6 configuration at $k=3$ (+8.3 points), and the largest advanced-task estimate is the same configuration at $k=10$ (+8.6 points). These descriptive differences do not identify difficulty as a causal moderator; Table~\ref{tab:difficulty-baselines} and Appendix Figure~\ref{fig:difficulty-deltas} report the full breakdown.

Task-level transitions and per-goal decomposition in Appendix~\ref{app:secondary-results} localize the largest negative shifts to postconditions, which Section~\ref{sec:mechanism} analyzes as the main formalization bottleneck.

%% file: sections/05_mechanism_failure_analysis.tex
\subsection{Failure Analysis}

\label{sec:mechanism}

The failure patterns in Section~\ref{sec:results} are consistent with clarified task information being lost during Lean formalization. The elicitation agent can add reference-grounded assumptions, edge cases, and output constraints to the handoff, but the LLM formalizer must still translate that information into valid Lean preconditions and postconditions. Compile-conditional and strict pass@1 can therefore move in different directions, although compile-conditional comparisons remain diagnostic because treatment changes the subset of outputs that compile.

The per-goal decomposition localizes the largest negative changes to postconditions. For Sonnet 4.6 same-model, precondition completeness increases by 3 tasks at $k=3$ and 6 tasks at $k=10$. Those changes do not translate into strict pass@1 because postconditions deteriorate: postcondition soundness falls by 13 tasks at $k=3$ and 19 tasks at $k=10$, while postcondition completeness falls by 7 and 15 tasks. Claude Sonnet 5 shows a related postcondition-sensitive pattern at $k=10$, losing 6 tasks on postcondition soundness and 11 tasks on postcondition completeness. Sonnet 4.6 to GPT-5.6 and Sonnet 4.6 to Gemini 3.8 Flash avoid large postcondition losses.

Overall, the failure analysis shows that final Lean results depend on whether the complete pipeline expresses the reference-grounded clarification in compilable and semantically correct specifications. Postconditions have the largest observed degradations; failures there can erase positive changes elsewhere. The qualitative cases reveal two distinct strict-score mechanisms: some transitions replace a semantically incorrect specification, while others replace an evaluator-unknown formulation with one Verina can certify. Appendix Table~\ref{tab:failure-examples} and Section~\ref{app:case-studies} report both kinds of evidence. Taken together, elicitation is configuration-dependent rather than uniformly beneficial: positive strict estimates occur in the Sonnet 4.6-to-GPT-5.6 and Sonnet 4.6-to-Gemini pipelines, while postcondition degradation offsets potential gains in other configurations.

%% file: sections/07_discussion.tex
\section{Discussion}

The results support a pipeline-level rather than formalizer-level interpretation. Clarification can help, but its effect depends on both the elicited handoff and its translation into Lean. Because every cross-model configuration independently samples its dialogue and handoff, those comparisons do not isolate the receiving formalizer or rank formalizers on a shared specification. Compile-conditional estimates are likewise diagnostic because baseline and treatment condition on different compiling subsets, and differences across Verina's difficulty labels do not establish difficulty or underspecification as a causal moderator.

Postconditions are the clearest bottleneck: useful task information can be lost when a richer handoff becomes an incomplete, over-specific, or otherwise incorrect Lean postcondition. The qualitative cases also distinguish semantic repair from evaluator certification: ComputeAvg has a concrete baseline counterexample, whereas the LongestCommonPrefix and RunLengthEncode baselines fail strict scoring only because some Verina checks return \texttt{UNKNOWN}. Interactive-autoformalization evaluations should therefore separate information acquired during elicitation, its translation into a formal artifact, and the evaluator's ability to certify that artifact. These results concern the controlled pipeline and do not establish performance with human developers.

%% file: sections/08_limitations.tex
\section{Limitations}
\label{sec:limitations}

The study answers the two research questions within a controlled benchmark setting, not a deployed human-facing workflow. Verina's task distribution and Lean idioms may not represent other verification settings or industrial workflows \citep{ye2026verina}. The reference-grounded simulated developer may also be more precise and consistent than an ordinary developer. Robustness to an imperfect answer source should be evaluated by systematically masking reference fields and injecting omissions, uncertainty, or inconsistent answers, then measuring how performance degrades. Transfer to human intent elicitation additionally requires real-developer studies with interface, trust, and accessibility evaluation \citep{trinh2026hilbench,raghavendra2026sweinteract}.

The experiment does not compare EA-selected questions against another question-selection policy using the same simulated developer and question budget. Consequently, it measures the value of the complete EA--simulated-developer interaction, not whether this EA selects better questions than fixed or generic clarification strategies. A question-policy control would hold the answer source and interaction budget fixed while varying only which questions are asked.

The question budgets are upper bounds, and the EA often stops before exhausting them. Because the realized number of questions depends on the task-specific state produced during elicitation, comparisons between $k=3$ and $k=10$ do not identify a causal effect of dialogue length. A fixed-turn experiment or a design that randomizes continuation after each exchange would be needed to isolate that effect.

The primary experiments use only Sonnet 4.6 as the EA and simulated developer. Moreover, each cross-model arm independently samples its dialogue and refined description, so the arms differ in both handoff and receiving formalizer; the auxiliary all-GPT-5.6 run changes every role. The results therefore compare complete configurations rather than isolating role-specific effects. A clean study would reuse saved handoffs across formalizers and separately evaluate non-Claude EAs and simulated developers under the same protocol.

The number of tested models is too small to support provider- or training-lineage claims. Evaluating model-role effects requires more Claude and non-Claude formalizers, non-Claude EAs, reverse cross-model pairings, and shared-handoff interventions that constrain postcondition structure.

The evaluation is limited to specification generation and does not test downstream proof or code generation \citep{ye2026verina}. It does not independently score the refined descriptions, and Verina treats \texttt{UNKNOWN} as strict failure; gains may therefore reflect either semantic correction or more certifiable reformulation. The qualitative examples distinguish these mechanisms but do not estimate their prevalence. Uncertainty intervals capture task-level variation, not repeated-generation variance, and the experiments do not establish the most cost-effective allocation of models across roles. The work could reduce specification mismatch, but deploying it without human review could also create false confidence when the answer source or formalizer is wrong; verifier checks mitigate but do not eliminate that risk.

%% file: sections/09_conclusion.tex
\section{Conclusion}

This paper studied whether an elicitation agent can obtain useful information from a reference-grounded simulated developer before Lean formalization. Reference-grounded clarification has positive verifier-checked point estimates in some complete pipeline configurations, but not uniformly; postconditions are the largest source of degradation. Because dialogues are resampled across configurations, the study does not isolate formalizer effects, and it does not establish performance with imperfect or human developers. Future work should compare question-selection policies, reuse shared handoffs to isolate model roles, vary the EA and answer source, and test human clarification.

%% file: sections/10_appendix.tex
\section{Appendix}
\label{sec:appendix}

\subsection{Secondary Results}
\label{app:secondary-results}

\begin{center}
\refstepcounter{table}\label{tab:mechanism}
\begin{minipage}{\textwidth}
\textbf{Table~\thetable:} Secondary evaluations aligned with Section~\ref{sec:methods}; reported in the appendix as secondary support for the difficulty split discussed in Section~\ref{sec:results}. Difficulty columns report strict pass@1 percentage-point changes for basic and advanced tasks. Gain/reg. reports task-level strict-pass gains and regressions. Pre and post columns report soundness/completeness changes under strict scoring, with the denominator fixed at 189 tasks.
\begin{center}
\resizebox{\textwidth}{!}{%
\begin{tabular}{llrrrrrr}
\toprule
Setting & Budget & Basic & Advanced & Gain/reg. & Pre S/C & Post S/C \\
\midrule
Sonnet 4.6 same-model & $k=3$ & -7.4 & +3.7 & 24/29 & -3/+3 & -13/-7 \\
Sonnet 4.6 same-model & $k=10$ & -8.3 & +1.2 & 26/34 & -6/+6 & -19/-15 \\
Sonnet 4.6 to GPT-5.6 & $k=3$ & +8.3 & +3.7 & 24/12 & +4/+3 & +9/+6 \\
Sonnet 4.6 to GPT-5.6 & $k=10$ & +4.6 & +8.6 & 25/13 & +4/+2 & +4/+6 \\
Sonnet 4.6 to Gemini 3.8 Flash & $k=3$ & +3.7 & +6.2 & 17/8 & +4/+1 & +3/+4 \\
Sonnet 4.6 to Gemini 3.8 Flash & $k=10$ & +5.6 & +3.7 & 20/11 & +2/+4 & +3/+3 \\
Sonnet 4.6 to Claude Sonnet 5 & $k=3$ & +0.0 & +1.2 & 19/18 & +1/+4 & +1/-7 \\
Sonnet 4.6 to Claude Sonnet 5 & $k=10$ & +3.7 & -12.3 & 24/30 & -1/+5 & -6/-11 \\
\bottomrule
\end{tabular}%
}
\end{center}
\end{minipage}
\end{center}

\begin{center}
\refstepcounter{table}\label{tab:realized-questions}
\begin{minipage}{\textwidth}
\textbf{Table~\thetable:} Realized question counts in the primary elicitation configurations. Mean, median, and range are computed over 189 task-level dialogues. Exhausted reports dialogues that used the full maximum budget. Because the EA can stop before reaching the cap, $k$ does not denote a fixed dialogue length.
\begin{center}
\begin{tabular}{llrrrr}
\toprule
Configuration & Budget & Mean & Median & Range & Exhausted \\
\midrule
Sonnet 4.6 same-model & $k=3$ & 2.58 & 3 & 1--3 & 144/189 (76.2\%) \\
Sonnet 4.6 same-model & $k=10$ & 7.03 & 7 & 1--10 & 48/189 (25.4\%) \\
Sonnet 4.6 to GPT-5.6 & $k=3$ & 2.61 & 3 & 1--3 & 145/189 (76.7\%) \\
Sonnet 4.6 to GPT-5.6 & $k=10$ & 6.93 & 8 & 1--10 & 39/189 (20.6\%) \\
Sonnet 4.6 to Gemini 3.8 Flash & $k=3$ & 2.67 & 3 & 1--3 & 152/189 (80.4\%) \\
Sonnet 4.6 to Gemini 3.8 Flash & $k=10$ & 7.17 & 8 & 1--10 & 43/189 (22.8\%) \\
Sonnet 4.6 to Claude Sonnet 5 & $k=3$ & 2.61 & 3 & 1--3 & 147/189 (77.8\%) \\
Sonnet 4.6 to Claude Sonnet 5 & $k=10$ & 7.07 & 7 & 1--10 & 40/189 (21.2\%) \\
\bottomrule
\end{tabular}
\end{center}
\end{minipage}
\end{center}

\begin{center}
\refstepcounter{table}\label{tab:strict-uncertainty}
\begin{minipage}{\textwidth}
\textbf{Table~\thetable:} Paired uncertainty estimates for strict pass@1 deltas. Gains/regressions count discordant task pairs. The p-value is an exact two-sided sign test, equivalent to a McNemar test on discordant pairs. Confidence intervals are paired task-level bootstrap 95\% intervals for the treatment-minus-baseline percentage-point delta.
\begin{center}
\resizebox{\textwidth}{!}{%
\begin{tabular}{llrrrr}
\toprule
Setting & Budget & Strict \tri{} (pp) & Gain/reg. & Sign/McNemar $p$ & Bootstrap 95\% CI (pp) \\
\midrule
Sonnet 4.6 same-model & $k=3$ & -2.6 & 24/29 & 0.583 & [-10.1, +4.8] \\
Sonnet 4.6 same-model & $k=10$ & -4.2 & 26/34 & 0.366 & [-12.2, +3.7] \\
Sonnet 4.6 to GPT-5.6 & $k=3$ & +6.3 & 24/12 & 0.065 & [+0.0, +12.7] \\
Sonnet 4.6 to GPT-5.6 & $k=10$ & +6.3 & 25/13 & 0.073 & [+0.0, +12.7] \\
Sonnet 4.6 to Gemini 3.8 Flash & $k=3$ & +4.8 & 17/8 & 0.108 & [-0.5, +10.1] \\
Sonnet 4.6 to Gemini 3.8 Flash & $k=10$ & +4.8 & 20/11 & 0.150 & [-1.1, +10.6] \\
Sonnet 4.6 to Claude Sonnet 5 & $k=3$ & +0.5 & 19/18 & 1.000 & [-5.8, +6.9] \\
Sonnet 4.6 to Claude Sonnet 5 & $k=10$ & -3.2 & 24/30 & 0.497 & [-10.6, +4.2] \\
\bottomrule
\end{tabular}%
}
\end{center}
\end{minipage}
\end{center}

\begin{center}
\refstepcounter{table}\label{tab:compile-uncertainty}
\begin{minipage}{\textwidth}
\textbf{Table~\thetable:} Bootstrap uncertainty estimates for compile-conditional pass@1 deltas. Baseline and treatment columns report strict passes among generated specifications that compile. Because the compiling set can change across conditions, these intervals are diagnostic and are not paired McNemar tests.
\begin{center}
\resizebox{\textwidth}{!}{%
\begin{tabular}{llrrrr}
\toprule
Setting & Budget & Compile-cond. \tri{} (pp) & Baseline & Treatment & Bootstrap 95\% CI (pp) \\
\midrule
Sonnet 4.6 same-model & $k=3$ & +3.8 & 114/162 & 109/147 & [-3.4, +11.2] \\
Sonnet 4.6 same-model & $k=10$ & +2.7 & 114/162 & 106/145 & [-5.3, +10.8] \\
Sonnet 4.6 to GPT-5.6 & $k=3$ & +4.6 & 112/165 & 124/171 & [-2.1, +11.4] \\
Sonnet 4.6 to GPT-5.6 & $k=10$ & +3.8 & 112/165 & 124/173 & [-3.1, +10.6] \\
Sonnet 4.6 to Gemini 3.8 Flash & $k=3$ & +4.9 & 113/183 & 122/183 & [-0.5, +10.6] \\
Sonnet 4.6 to Gemini 3.8 Flash & $k=10$ & +4.9 & 113/183 & 122/183 & [-1.1, +11.0] \\
Sonnet 4.6 to Claude Sonnet 5 & $k=3$ & -1.9 & 120/167 & 121/173 & [-8.0, +4.3] \\
Sonnet 4.6 to Claude Sonnet 5 & $k=10$ & -2.3 & 120/167 & 114/164 & [-9.4, +4.6] \\
\bottomrule
\end{tabular}%
}
\end{center}
\end{minipage}
\end{center}

\begin{center}
\refstepcounter{table}\label{tab:failure-examples}
\begin{minipage}{\textwidth}
\textbf{Table~\thetable:} Paired strict-pass gain and regression examples for the three evaluator-grounded failure categories. All examples are from the Sonnet 4.6 to GPT-5.6, $k=3$ condition. Categories can overlap when noncompilation prevents semantic scoring; the soundness and completeness rows isolate changes in the named semantic criterion while both artifacts compile.
\smallskip
\begin{tabularx}{\textwidth}{@{}lXX@{}}
\toprule
Category & Gain example & Regression example \\
\midrule
Compilation & \texttt{verina\_basic\_103}: the baseline postcondition does not compile; the treatment's relational postcondition compiles and passes all checks. & \texttt{verina\_basic\_23}: the baseline compiles and passes; the treatment's helper-based precondition and postcondition fail to compile. \\
Soundness & \texttt{verina\_basic\_107}: replacing an interval approximation with exact integer division removes the counterexample $a=1$, $b=0$, $result=1$ and changes postcondition soundness from fail to pass. & \texttt{verina\_basic\_34}: replacing exact filtering with membership and sublist clauses changes postcondition soundness from pass to fail. \\
Completeness & \texttt{verina\_basic\_10}: adding the required nonempty-array precondition changes precondition completeness from fail to pass. & \texttt{verina\_basic\_32}: an expanded endpoint-and-middle-element postcondition changes postcondition completeness from pass to unknown. \\
\bottomrule
\end{tabularx}
\end{minipage}
\end{center}

\subsection{Appendix Figures}
\label{app:figures}

\begin{center}
\refstepcounter{figure}\label{fig:ea-architecture}
\includegraphics[width=0.98\linewidth]{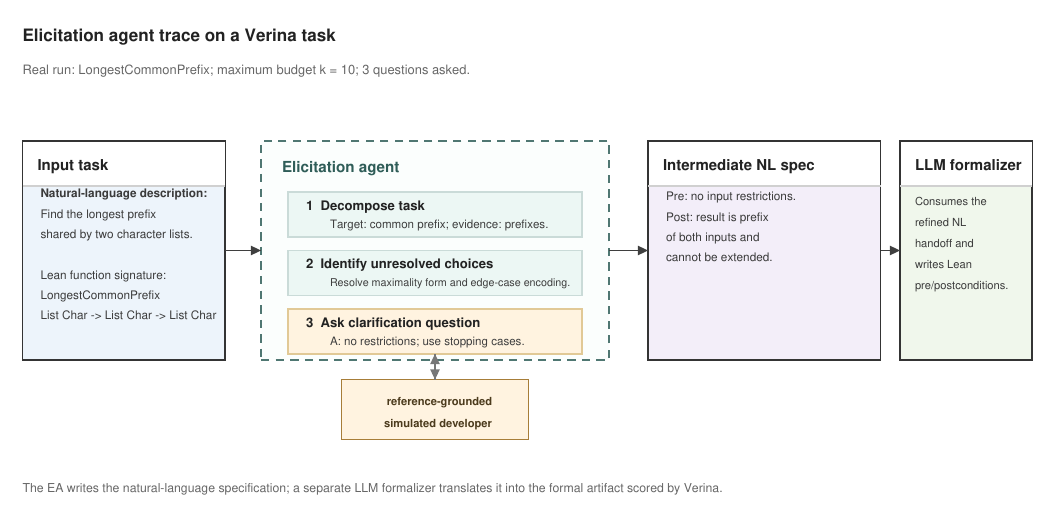}
\vspace{0.2em}
\begin{minipage}{0.94\linewidth}
\small\textbf{Figure~\thefigure:} Elicitation-agent components illustrated with shortened outputs from a real Verina treatment run. The agent receives an informal task description and Lean function signature, decomposes the task, identifies unresolved choices, asks bounded clarification questions, and writes an intermediate natural-language specification. This run asked three questions under a maximum budget of ten before a separate LLM formalizer wrote Lean.
\end{minipage}
\end{center}

\begin{center}
\refstepcounter{figure}\label{fig:result-deltas}
\includegraphics[width=\linewidth]{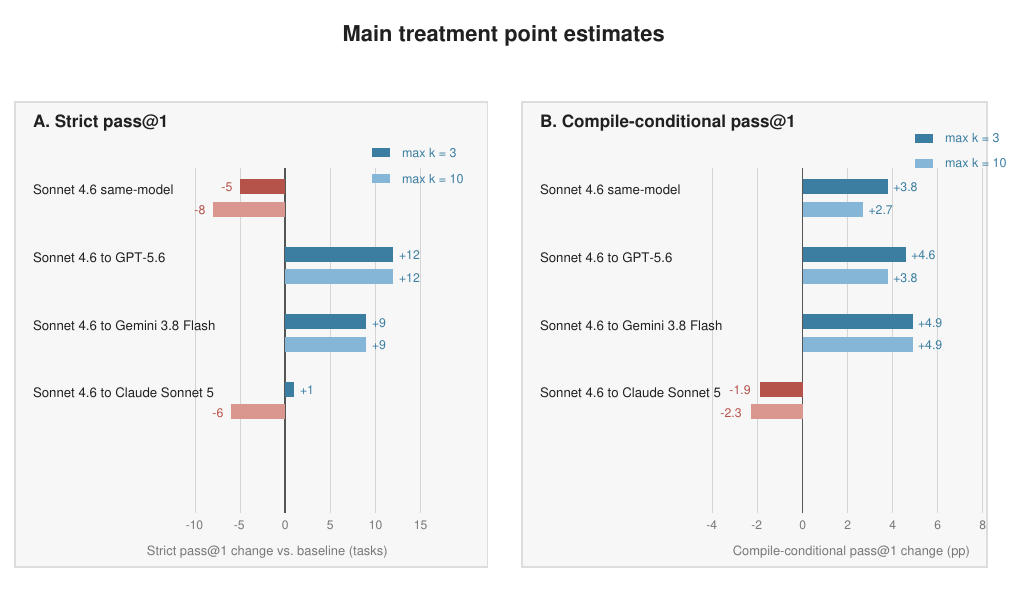}
\vspace{0.3em}
\begin{minipage}{0.94\linewidth}
\textbf{Figure~\thefigure:} Main treatment point estimates. Here $k$ is the maximum number of questions, not the realized dialogue length. Rows name complete configurations; ``to'' indicates Claude Sonnet 4.6 as elicitation agent and reference-grounded simulated developer with the target model as LLM formalizer. The EA--simulated-developer interaction was sampled independently for each configuration, so the figure does not rank formalizers on a shared elicited specification. The Claude Sonnet 5 configuration has negative compile-conditional changes at both budgets and a strict pass@1 regression at $k=10$.
\end{minipage}
\end{center}

\begin{center}
\refstepcounter{figure}\label{fig:difficulty-deltas}
\includegraphics[width=\linewidth]{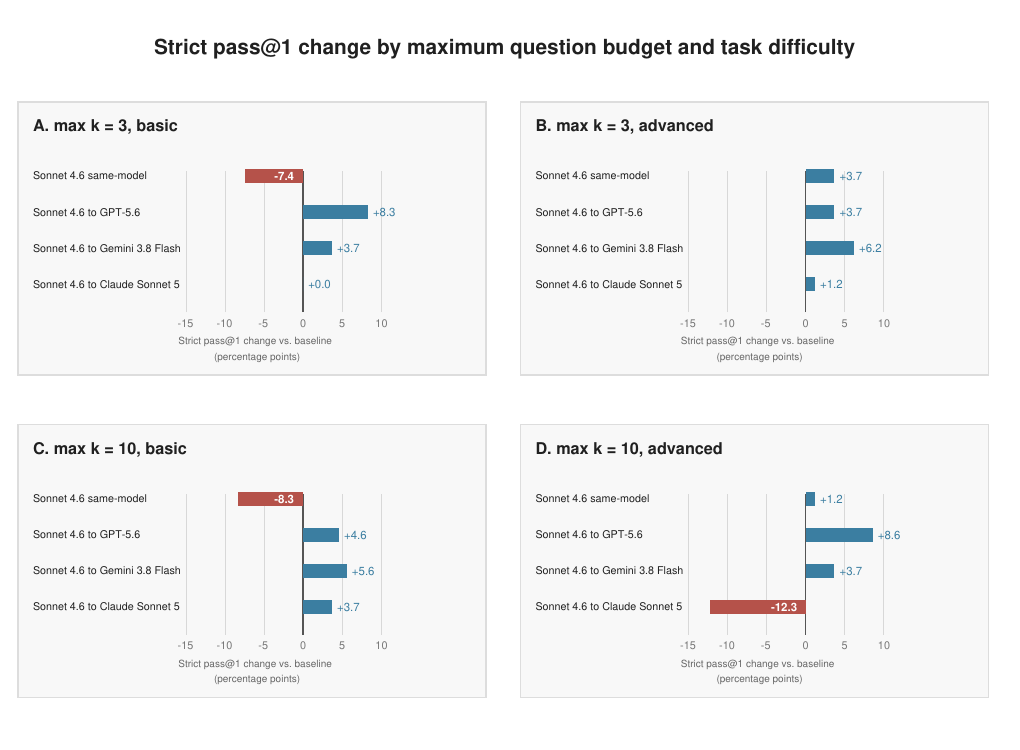}
\vspace{0.3em}
\begin{minipage}{0.94\linewidth}
\textbf{Figure~\thefigure:} Strict pass@1 point estimates by maximum question budget and Verina difficulty split. ``To'' rows are independently sampled cross-model configurations, not formalizer-only ablations. Sonnet 4.6 same-model loses on basic tasks at both budgets, while Claude Sonnet 5 regresses on advanced tasks at $k=10$. Table~\ref{tab:mechanism} reports the full secondary breakdown.
\end{minipage}
\end{center}

\begin{center}
\refstepcounter{figure}\label{fig:postcondition-deltas}
\includegraphics[width=\linewidth]{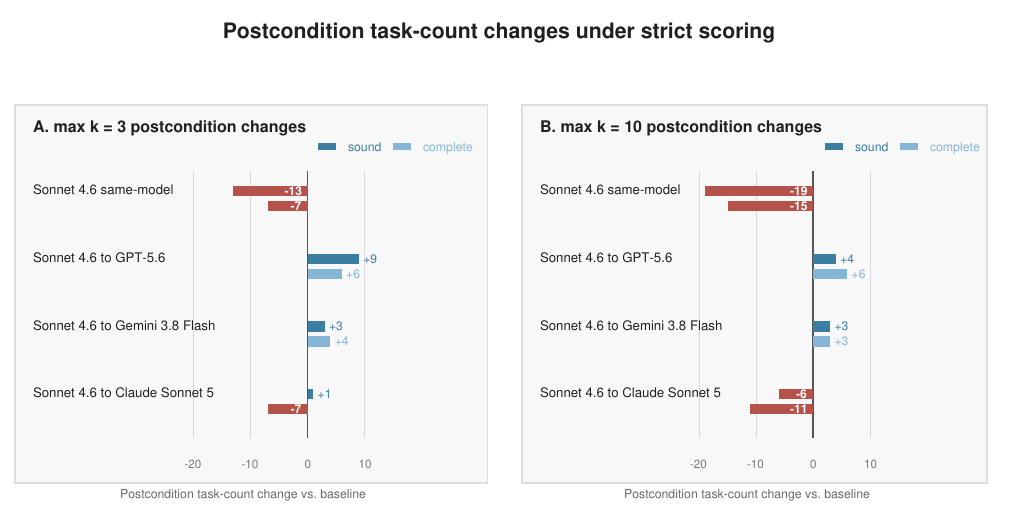}
\vspace{0.3em}
\begin{minipage}{0.94\linewidth}
\textbf{Figure~\thefigure:} Postcondition changes in task counts under strict scoring. ``To'' rows are independently sampled cross-model configurations. Sonnet 4.6 same-model has large negative postcondition soundness and completeness shifts at both budgets, and the Claude Sonnet 5 configuration shows a related pattern at $k=10$. The GPT-5.6 and Gemini 3.8 Flash configurations avoid the large postcondition collapse.
\end{minipage}
\end{center}

\subsection{Raw Example Trace for Figure~\ref{fig:ea-architecture}}
\label{app:figure1-example}

Figure~\ref{fig:ea-architecture} summarizes a treatment run for \emph{verina\_basic\_71} from the \emph{gemini\_mixed\_full\_k10} condition, where Claude Sonnet 4.6 served as the elicitation agent and reference-grounded simulated developer, and Gemini 3.8 Flash served as the LLM formalizer. The example illustrates the implemented reference-grounded clarification condition. The simulated developer answers in natural language and does not emit Lean code, while still providing detailed contract structure for the downstream LLM formalizer. The blocks below contain raw run text, with only LaTeX escaping and line wrapping applied.
\par\medskip\noindent\textbf{Illustrative certification gain: longest common prefix.}\par\nobreak\smallskip
Baseline: strict fail because some postcondition checks are \texttt{UNKNOWN}. Treatment: strict pass.
\par\medskip\noindent\textbf{Raw original input.}\par\nobreak\smallskip
\begin{lstlisting}[style=appendixblock]
Function signature: LongestCommonPrefix(str1 : List Char, str2 : List Char) ->
List Char
Informal description: -----Description----- This problem involves determining the
longest common prefix shared by two lists of characters. Given two sequences, the
goal is to identify and return the maximal contiguous sequence of characters from
the beginning of both lists that are identical.
-----Input----- The input consists of: -- str1: A list of characters. -- str2: A
list of characters.
-----Output----- The output is a list of characters representing the longest
common prefix of the two input lists. The output list satisfies the following
conditions: -- Its length is less than or equal to the length of each input list.
-- It is exactly the prefix of both str1 and str2. -- It is empty if the first
characters of the inputs differ or if one of the lists is empty.
-----Note----- It is assumed that both inputs are provided as valid lists of
characters. The function always returns the correct longest common prefix based on
the inputs.
\end{lstlisting}
\medskip\noindent\textbf{Raw clarification dialogue.}\par\nobreak\smallskip
\begin{lstlisting}[style=appendixblock]
(*@\textnormal{\textbf{Turn 1 elicitation agent:}}@*) What should the precondition be? Is it literally 'True', should it be omitted entirely, or should it assert something about the inputs even if that something is trivially satisfied?
(*@\textnormal{\textbf{Turn 1 simulated developer:}}@*) It's literally just True --- there are no restrictions on the inputs. Both strings can be empty, identical, completely different, or one can be a prefix of the other; all cases are valid.

(*@\textnormal{\textbf{Turn 2 elicitation agent:}}@*) Should 'longest' be expressed as a universal prefix-order condition (every common prefix is a prefix of the result) or as a length-based condition (no common prefix is strictly longer)?
(*@\textnormal{\textbf{Turn 2 simulated developer:}}@*) Length-based is fine --- the result should be as long as possible, meaning there's no longer string that is also a common prefix of both inputs.

(*@\textnormal{\textbf{Turn 3 elicitation agent:}}@*) How many conjuncts should the postcondition have, and what is each one? Should the empty-list edge case be a separate conjunct or left as a corollary of the maximality clause?
(*@\textnormal{\textbf{Turn 3 simulated developer:}}@*) There are five conjuncts: the result's length doesn't exceed str1's length, the result equals the first that-many characters of str1, the result's length doesn't exceed str2's length, the result equals the first that-many characters of str2, and a maximality clause saying either we consumed all of str1, or all of str2, or the next characters in str1 and str2 differ. The empty-list edge case falls out naturally from the maximality clause --- no separate conjunct needed.
\end{lstlisting}
\medskip\noindent\textbf{Raw refined natural-language specification.}\par\nobreak\smallskip
\begin{lstlisting}[style=appendixblock]
# Refined Description: LongestCommonPrefix
## Overview
This function computes the longest common prefix of two lists of characters. Given
two character lists, it identifies and returns the maximal contiguous sequence of
characters from the beginning of both lists that are identical, proceeding
character by character from the front until one list is exhausted or the next
characters disagree.
## Precondition
The function accepts any two lists of characters, str1 and str2, without
restriction. Both inputs are always valid because a list of characters in Lean 4
is an inductive type with no invalid inhabitants. Either or both inputs may be
empty, identical, completely disjoint, or one may be a prefix of the other. The
precondition is trivially true and imposes no constraints beyond the inputs being
well-typed lists of characters.
## Postcondition
Let k denote the length of the returned result. The result must satisfy all five
of the following conditions simultaneously.
First, k is less than or equal to the length of str1. Second, the result equals
exactly the first k characters of str1, that is, the result is the prefix of str1
of length k. Third, k is less than or equal to the length of str2. Fourth, the
result equals exactly the first k characters of str2, that is, the result is the
prefix of str2 of length k. Fifth, the result is maximal in the sense that the
scan could not have been extended further: either k equals the length of str1
(str1 is fully consumed), or k equals the length of str2 (str2 is fully consumed),
or the character at position k in str1 differs from the character at position k in
str2 (the next characters diverge). This fifth condition is expressed as a
disjunction of these three stopping cases.
Conjuncts one and three are load-bearing length bounds and must not be omitted;
they are not derivable from any other single conjunct in this formulation.
Conjuncts two and four together with the length bounds imply that the result is a
prefix of both str1 and str2 in the standard sense that each input equals the
result concatenated with some suffix. The fifth conjunct encodes the termination
condition of a character-by-character scan and directly mirrors the structure of a
recursive implementation, making it easier to verify from such an implementation
than a universally quantified maximality clause would be.
The empty-list edge case requires no separate clause. When str1 is empty, the
first length bound forces k to be zero, and the fifth conjunct is satisfied by its
first disjunct. When str2 is empty, the same reasoning applies via the third
length bound and the second disjunct. When the first characters of the two lists
differ, k is zero and the fifth conjunct is satisfied by its third disjunct. All
of these are corollaries of the five-conjunct postcondition taken together.
The result is uniquely determined by the postcondition. If two lists p and q both
satisfied all five conjuncts, a comparison of their lengths would show they must
be equal: if one were strictly shorter than the other, the shorter one's fifth
conjunct would require the next characters to disagree or a list to be exhausted,
but the longer solution's existence as a common prefix of both lists at that
position would contradict that requirement. Once the lengths are shown equal,
conjuncts two and four force the contents to agree.
## Notes on Formulation Choices
The fifth conjunct uses a disjunctive stopping condition rather than a universal
quantifier over all common prefixes. This disjunctive form is logically
equivalent, in the context of the full five-conjunct postcondition, to both the
prefix-order universal form stating that every common prefix of str1 and str2 is
itself a prefix of the result, and the length-based universal form stating that
every common prefix has length at most k. The disjunctive form is the canonical
formulation chosen by the developer because it avoids universal quantification and
is more directly amenable to proof by induction on the shared prefix length.
Downstream proofs that require the universal form can derive it from the fifth
conjunct via the equivalence between these formulations.
The postcondition is expressed using list length, list prefix extraction by
length, and character lookup by index. No Mathlib import is strictly required for
stating the postcondition, though Mathlib lemmas connecting prefix extraction to
the standard list prefix relation may be useful in proofs. Character equality
appears only in the third disjunct of the fifth conjunct, where it is
propositional equality of optional characters; the none cases at that position are
ruled out in context by the length bounds and the exhaustion disjuncts, so no case
split on optionality is required in the statement itself.
The developer confirmed that the precondition is literally true with no
restrictions, that the five-conjunct postcondition is the intended specification,
and that the disjunctive stopping condition is the canonical maximality clause.
The developer had not previously considered the question of whether uniqueness
should be stated as a companion lemma or left implicit; the downstream generator
should treat uniqueness as a derivable consequence of the postcondition and may
optionally produce a companion lemma asserting that any two lists satisfying all
five conjuncts with respect to the same inputs must be equal.
\end{lstlisting}
\medskip\noindent\textbf{Generated Lean/spec contrast.}\par\nobreak\smallskip
\begin{lstlisting}[style=appendixblock]
Baseline generated precondition: True
Baseline generated postcondition: result.isPrefixOf str1 = true and
result.isPrefixOf str2 = true and forall p : List Char, p.isPrefixOf str1 = true
-> p.isPrefixOf str2 = true -> p.length <= result.length
Treatment generated precondition: True
Treatment generated postcondition: result.length <= str1.length and result =
str1.take result.length and result.length <= str2.length and result = str2.take
result.length and (result.length = str1.length or result.length = str2.length or
str1[result.length]? != str2[result.length]?)
\end{lstlisting}
\medskip\noindent\textbf{Exact Verina outcome.}
Both generated specifications compile, and all precondition checks pass. For the baseline postcondition, 9 of 11 soundness checks pass and 2 are \texttt{UNKNOWN}; 1 of 5 completeness checks passes and 4 are \texttt{UNKNOWN}. No check explicitly fails. The treatment postcondition passes all 11 soundness and all 5 completeness checks. Thus, this strict-score transition is a Verina certification gain: the treatment replaces an uncertified formulation with one the evaluator can certify, but the recorded verdict does not provide a semantic counterexample to the baseline.
\subsection{Failure-Analysis Case Studies}
\label{app:case-studies}

The LongestCommonPrefix and RunLengthEncode traces are strict-score certification gains under reference-grounded clarification: their baseline formulations compile and have no explicit failing check, but Verina returns \texttt{UNKNOWN} on some checks. ComputeAvg is a semantic repair with a concrete baseline counterexample. Together, the cases distinguish reformulation into an evaluator-certifiable specification from correction of demonstrated semantic error. They do not establish that a human developer would provide the same information.
\par\medskip\noindent\textbf{Advanced certification gain: run-length-encoding postconditions.}\par\nobreak\smallskip
Baseline: strict fail because some postcondition checks are \texttt{UNKNOWN}. Treatment: strict pass.
\par\medskip\noindent\textbf{Raw original input.}\par\nobreak\smallskip
\begin{lstlisting}[style=appendixblock]
Function signature: runLengthEncode(s : String) -> List (Char x Nat)
Informal description: -----Description----- This task requires writing a Lean 4
method that performs run-length encoding on a given string. The method should scan
the string from left to right and group consecutive identical characters into
pairs. Each pair consists of the character itself and the number of times it
appears consecutively. For example, "aaabbc" becomes [('a', 3), ('b', 2), ('c',
1)].
The resulting encoded list must satisfy the following properties: 1. No pair has a
zero or negative run-length. 2. Consecutive pairs in the encoding list must not
have the same character. 3. Decoding the output should return the original string.
-----Input----- The input is a single string, `s`.
-----Output----- The output is a list of pairs `(Char, Nat)`, which represents the
run-length-encoded form of the input string.
\end{lstlisting}
\medskip\noindent\textbf{Raw clarification dialogue.}\par\nobreak\smallskip
\begin{lstlisting}[style=appendixblock]
(*@\textnormal{\textbf{Turn 1 elicitation agent:}}@*) Should the postcondition characterise run-length encoding by listing all three properties independently, or should one property (e.g., faithful decoding) be treated as the primary condition with the others as supporting invariants?
(*@\textnormal{\textbf{Turn 1 simulated developer:}}@*) All three properties are equally important and should stand independently: every run has a positive count, no two adjacent runs share the same character, and decoding the result gives back the original string.

(*@\textnormal{\textbf{Turn 2 elicitation agent:}}@*) How should the decode inverse function be defined for use inside the postcondition --- as an inline expression, a named auxiliary, or an imported standard definition?
(*@\textnormal{\textbf{Turn 2 simulated developer:}}@*) A named auxiliary function defined just before the postcondition is fine --- something that takes the run-length encoded list and reconstructs the original string by repeating each character by its count and concatenating everything together.

(*@\textnormal{\textbf{Turn 3 elicitation agent:}}@*) How should the 'no two consecutive pairs share the same character' invariant be formally expressed --- as a universally quantified index condition, a pairwise predicate on list windows, or a recursive structural predicate?
(*@\textnormal{\textbf{Turn 3 simulated developer:}}@*) A universally quantified index condition is fine --- for every valid index i, the character at position i should differ from the character at position i+1. Either of the other formulations would also be acceptable if they're equivalent.
\end{lstlisting}
\medskip\noindent\textbf{Raw refined natural-language specification.}\par\nobreak\smallskip
\begin{lstlisting}[style=appendixblock]
# Refined Description: Run-Length Encoding (`runLengthEncode`)
## Summary
`runLengthEncode` takes a single string and produces its canonical run-length
encoding as a list of character-count pairs. Scanning the string from left to
right, it groups every maximal contiguous run of identical characters into one
pair consisting of the character and the number of times it appears consecutively
in that run. For example, the string "aaabbc" becomes the list containing the pair
('a', 3), followed by ('b', 2), followed by ('c', 1). The function applies to all
valid Lean 4 strings, including the empty string and strings containing any
Unicode scalar value such as emoji, accented letters, or CJK characters.
## Precondition
The function imposes no restrictions on its input. It is total and terminates for
every value of type String, including the empty string and strings containing
arbitrary Unicode scalar values. No non-emptiness requirement, no ASCII
restriction, and no other constraint on the input is assumed. The precondition is
simply true and may be omitted from any formal theorem statement.
## Postcondition
Let `output` denote the list returned by `runLengthEncode s`. The output must
satisfy all three of the following properties simultaneously, as co-equal
conjuncts with no single property designated primary.
First, every run-length in the output is strictly positive: for every pair in
`output`, the count component is greater than zero. Because Lean 4's `Nat` type is
non-negative by construction, this reduces to requiring that no count is zero;
negative counts are impossible by type.
Second, no two consecutive pairs in the output share the same character: for every
valid index i into `output`, the character at position i differs from the
character at position i plus one. This condition is equivalent to a recursive
structural check that each adjacent pair of entries has distinct first components,
and it is also equivalent to the built-in `List.Chain'` predicate applied with
character inequality. Any of these equivalent formulations is acceptable. This
condition implicitly enforces maximality of runs: because no two adjacent pairs
may share a character, a run of five identical characters cannot be split into two
adjacent pairs, and the greedy left-to-right maximal grouping is the unique
encoding that satisfies all three conditions.
Third, decoding the output recovers the original string exactly. Decoding is
performed by a named auxiliary function, here called `rleDecodeToString`, defined
as follows: given the list of pairs, for each pair replicate the character by its
count to produce a list of characters, concatenate all such lists in order, and
convert the resulting character list back to a string using `String.mk`. Formally,
`rleDecodeToString pairs` equals `String.mk` applied to the result of binding each
pair `(c, n)` through `List.replicate n c`. The postcondition requires that
`rleDecodeToString output` equals `s`. This definition is a named, standalone
auxiliary whose correctness is part of the trusted specification base, not derived
from a library import. Character equality throughout uses Lean 4's default
`DecidableEq Char`, which is Unicode scalar value equality, case-sensitive, with
no normalisation.
These three conditions together are jointly sufficient to characterise the
canonical run-length encoding uniquely. A list such as `[('a', 3), ('a', 1)]`
would satisfy the decode round-trip for the input "aaaa" but would violate the
no-adjacent-duplicate condition, confirming that all three conjuncts are
necessary. Left-to-right ordering of the output pairs is enforced implicitly by
the decode round-trip: any reordering of pairs would produce a decoded string
different from `s`.
## Edge Cases and Boundary Behavior
When the input is the empty string, the output must be the empty list. This is not
a separate conjunct but is derivable from the three main conditions: the decode of
the empty list is the empty string, and the positive-run and no-adjacent-duplicate
conditions hold vacuously over an empty list.
When the input is a single character such as "a", the output must be the
single-element list containing `('a', 1)`. The no-adjacent-duplicate condition
holds vacuously for a one-element list, the count 1 satisfies the positivity
condition, and the decode round-trip is straightforwardly satisfied. No
special-case conjunct is needed.
When the input consists entirely of one repeated character such as "aaaa", the
output must be the single-element list `[('a', 4)]`. Any multi-pair encoding such
as `[('a', 2), ('a', 2)]` is ruled out by the no-adjacent-duplicate condition.
This is a consequence of the existing conjuncts and requires no additional clause.
## Points the Developer Had Not Considered
No points arose in the dialogue where the developer said "I hadn't considered
that." All edge cases and design decisions were resolved by domain knowledge and
confirmed assumptions. The downstream generator should apply permissive defaults
only if it encounters a situation not covered above, and in that case should
prefer the minimal formulation that does not add conjuncts beyond the three
co-equal postcondition clauses described here.
\end{lstlisting}
\medskip\noindent\textbf{Generated Lean/spec contrast.}\par\nobreak\smallskip
\begin{lstlisting}[style=appendixblock]
Baseline generated precondition: True
Baseline generated postcondition: (forall run in result, 0 < run.2) and (forall i
: Nat, i + 1 < result.length -> (result[i]!).1 != (result[i + 1]!).1) and
result.flatMap (fun run => List.replicate run.2 run.1) = s.toList
Treatment generated precondition: True
Treatment generated postcondition: (forall p in result, 0 < p.2) and List.Chain'
(fun x y : Char x Nat => x.1 != y.1) result and rleDecodeToString result = s
\end{lstlisting}

\medskip\noindent\textbf{Exact Verina outcome.}
Both generated specifications compile, and all precondition checks pass. The baseline postcondition passes all 6 soundness checks; 2 of 5 completeness checks pass and 3 are \texttt{UNKNOWN}. No check explicitly fails. The treatment postcondition passes all 6 soundness and all 5 completeness checks. As with LongestCommonPrefix, the evidence supports a Verina certification gain rather than a counterexample-based semantic repair.

\medskip\noindent\textbf{Counterexample-based semantic gain: exact integer average.}\par\nobreak\smallskip
The \texttt{verina\_basic\_107} task, ComputeAvg, is a gain in the Sonnet 4.6-to-GPT-5.6 configuration at $k=3$. Both specifications compile and all precondition checks pass. The baseline postcondition is
\[
  a+b-1 \leq 2\,result \;\wedge\; 2\,result \leq a+b+1,
\]
which is an interval approximation rather than an exact integer average. For $a=1$ and $b=0$, it admits $result=1$ because $0 \leq 2 \leq 2$, although the intended result is $(1+0)/2=0$. Verina accordingly records 14 passing and 1 failing postcondition-soundness check; all 5 completeness checks pass.

The EA asked whether the postcondition should use a bracket inequality or direct equality to a specific division operation. The simulated developer requested exact integer-division behavior, and the refined handoff led the LLM formalizer to produce
\[
  result=(a+b)/2.
\]
The treatment passes all 15 postcondition-soundness and all 5 postcondition-completeness checks. Unlike the two certification examples above, this transition removes a concrete behavior admitted by the baseline but excluded by the intended specification.